# Microwave photonic fractional Hilbert transformer with an integrated optical soliton crystal micro-comb

Mengxi Tan, Xingyuan Xu, Bill Corcoran, Jiayang Wu, Andreas Boes, Thach G. Nguyen, Sai T. Chu, Brent E. Little, Roberto Morandotti, Arnan Mitchell, and David J. Moss

*Abstract*— **We report a photonic microwave and RF fractional Hilbert transformer based on an integrated Kerr micro-comb source. The micro-comb source has a free spectral range (FSR) of 50GHz, generating a large number of comb lines that serve as a high-performance multi-wavelength source for the transformer. By programming and shaping the comb lines according to calculated tap weights, we achieve both arbitrary fractional orders and a broad operation bandwidth. We experimentally characterize the RF amplitude and phase response for different fractional orders and perform system demonstrations of real-time fractional Hilbert transforms. We achieve a phase ripple of < 0.15 rad within the 3-dB pass-band, with bandwidths ranging from 5 to 9 octaves, depending on the order. The experimental results show good agreement with theory, confirming the effectiveness of our approach as a new way to implement high-performance fractional Hilbert transformers with broad processing bandwidth, high reconfigurability, and greatly reduced size and complexity.**

*Index Terms*—Kerr frequency comb, Hilbert transform, integrated optics, all-optical signal processing.

## I. INTRODUCTION

As one of the fundamental mathematical operations in modern signal processing, the Hilbert transform has found wide applications in radar systems, signal sideband modulators, measurement systems, speech processing, signal sampling, and many others [1]. While standard Hilbert transforms perform a ± 90° phase shift around a central frequency, with an all-pass amplitude transmission, fractional Hilbert transforms (FHTs) provide an additional degree of freedom in terms of a variable phase shift. They were introduced to realize versatile processing functions in order to meet specific requirements for practical applications such as hardware keys [2, 3], and they also play an important role in secure single sideband communications [2]. They also find applications in forming an image that is edge enhanced relative to the input object - the FHT was introduced to be able to select which edges are enhanced and to what degree the edge enhancement occurs [4].

Fractional Hilbert transformers based on electronics are subject to the intrinsic electronic bandwidth bottleneck [2, 5], thus significantly limiting their processing speed. In contrast, photonic technologies offer broad operation bandwidths as well as strong immunity to electromagnetic interference and thus are attractive for implementing high-speed fractional Hilbert transformers. Hilbert transformers based on free-space optics have been demonstrated [3, 6], which, although achieving attractive performance, are often bulky and suffer from significant complexity. Hilbert transformers based on phase-shifted fibre Bragg gratings can achieve broad operation bandwidths up to a few hundred gigahertz [7-10]. However, these methods can only yield precise FHTs for signals with specific bandwidths, for fixed fractional orders, and only produce the FHT of complex optical fields. This holds true for recent reports of an integrated reconfigurable microwave processor [11] as well as a simple, wideband and direct route to realize continuously tunable FHT based on a Bragg grating [12]. In practice, however, the FHT of the temporal intensity profiles associated with radio frequency (RF) and microwave signals, and not the complex optical fields, is required in many applications such as ultra-wideband frequency generation, RF measurement, and RF signal reshaping [13-16].

This work was supported by the Australian Research Council Discovery Projects Program (No. DP150104327). RM acknowledges support by the Natural Sciences and Engineering Research Council of Canada (NSERC) through the Strategic, Discovery and Acceleration Grants Schemes, by the MESI PSR-SIIRI Initiative in Quebec, and by the Canada Research Chair Program. He also acknowledges additional support by the Government of the Russian Federation through the ITMO Fellowship and Professorship Program (grant 074-U 01) and by the 1000 Talents Sichuan Program in China. Brent E. Little was supported by the Strategic Priority Research Program in China. Brent E. Little was supported by the Strategic Priority Research Program of the Chinese Academy of Sciences, Grant No. XDB24030000.

M. X. Tan, X. Y. Xu, J. Y. Wu, and D. J. Moss are with Centre for Micro-Photonics, Swinburne University of Technology, Hawthorn, VIC 3122, Australia.

A. Boes, T. G. Thach and A. Mitchell are with the School of Engineering, RMIT University, Melbourne, VIC 3001, Australia.

B. Corcoran is with Department of Electrical and Computer Systems Engineering, Monash University, Clayton, 3800 VIC, Australia.

S. T. Chu is with Department of Physics, City University of Hong Kong, Tat Chee Avenue, Hong Kong, China.

B. E. Little is with State Key Laboratory of Transient Optics and Photonics, Xi'an Institute of Optics and Precision Mechanics, Chinese Academy of Science, Xi'an, China.

R. Morandotti is with INRS-Énergie, Matériaux et Télécommunications, 1650 Boulevard Lionel-Boulet, Varennes, Québec, J3X 1S2, Canada. He is also a visiting professor with ITMO University, St. Petersburg, Russia, and with the Institute of Fundamental and Frontier Sciences, University of Electronic Science and Technology of China, Chengdu 610054, China.

In recent years, integrated fractional Hilbert transformers based on microring/microdisk resonators and Bragg gratings on a silicon-on-insulator (SOI) platform have been demonstrated, as well as an integrated InP-InGaAsP photonic chips, which can all provide competitive advantages such as compact device footprint, high stability, and mass-producibility [11,12, 17-19]. Better tunability has also been achieved by introducing thermo-optic microheaters to tune the phase shift. Nevertheless, these integrated devices can still only perform FHTs for complex optical fields and in most cases only provide a limited tuning range in the fractional order phase shift. In order to perform FHTs for the temporal intensity profiles of RF signals, other approaches based on transversal filter structures have been proposed, which can offer high reconfigurability by adjusting the tap coefficients, thus allowing precise control of the fractional order in a broad tuning range [20, 21]. One drawback to this approach, however, is that photonic transversal filters typically rely on multiple discrete laser sources that result in significantly increased system size, cost, and complexity, which limits the number of available taps and results in degraded processing performance. Therefore, instead of using individual light sources for each tap, a single source that can simultaneously generate a large number of high-quality wavelength channels would be highly desirable.

Optical frequency comb (OFC) technology has provided a powerful solution to address this issue and has attracted significant interests in recent years [22-36]. Various types of OFCs generated by mode-locked lasers [37], electro-optical modulators [38], and optical micro-resonators [39] have been demonstrated. Among them, Kerr OFCs generated by micro-resonators, or micro-combs, particularly those based on CMOS-compatible integrated platforms, can provide a large number of wavelength channels with greatly reduced footprint and complexity [40- 49].

Recently, [49] we demonstrated a photonic Hilbert transformer with up to 20 taps, based on a 200GHz spaced micro-comb source that achieved record performance in terms of RF bandwidth (5 octaves). Here, we demonstrate a fractional Hilbert transformer based on a Kerr micro-comb source with a significantly finer 50GHz comb spacing that provides a much larger number of comb lines (80) that serves as a high-performance multi-wavelength source for the photonic RF transversal filter. By programming and shaping the comb lines according to judiciously calculated tap weights, we experimentally demonstrate a FHT with arbitrary fractional orders and very broad operation bandwidths. We characterize the RF amplitude and phase response of the fractional Hilbert transformer with variable phase shifts of 15°, 30°, 45°, 60°, 75°, and 90° that correspond to tunable fractional orders of 0.17, 0.33, 0.5, 0.67, 0.83, and 1, respectively. We achieve operation over more than 5 octaves from 480MHz to 16.45GHz for the 90° phase-shift FHT and ±0.07 rad phase variation within the 3-dB bandwidth. System demonstrations of a real-time FHT for Gaussian pulse input signals are also performed. For the 0.166, 0.333, 0.5, 0.667, 0.833, and 1 fractional order Hilbert transformers, with root-mean-square errors (RMSEs) between the measured and theoretical curves being ~2.73%, ~2.75%,

~2.69%, ~2.69%, ~2.92%, ~2.85%, respectively. This good agreement with theory confirms our approach as an effective way to implement high-speed reconfigurable fractional Hilbert transformers with reduced footprint, lower complexity, and potentially lower cost.

## II. PRINCIPLE OF OPERATION

The spectral transfer function of a fractional Hilbert transformer can be expressed as [3, 50]:

$$H_P(\omega) = \begin{cases} e^{-j\varphi}, & if\ 0 \le \omega < \pi \\ e^{j\varphi}, & if\ -\pi \le \omega < 0 \end{cases} \quad (1)$$

where $\varphi = P \times \pi / 2$ is the phase shift, with $P$ denoting the fractional order. As can be seen from Eq. (1), a fractional Hilbert transformer can be regarded as a phase shifter with $\pm \varphi$ phase shifts around the centre frequency $\omega$ and it becomes a classical Hilbert transformer when $P = 1$. The corresponding impulse response is a continuous hyperbolic function given by [49, 50]:

$$h_P(t) = \begin{cases} \frac{1}{\pi t}, t \ne 0 \\ \cot(\varphi), t = 0 \end{cases} \quad (2)$$

The hyperbolic function is truncated and sampled in time by discrete taps. The sample spacing $\Delta t$ determines the null frequency $f_c = 1/\Delta t$. The order of the fractional Hilbert transformer is continuously tunable by only adjusting the coefficient of the $9^{th}$ ($t = 0$) tap while keeping the same coefficients for the other taps [3]. The theoretical amplitude and phase response of a fractional Hilbert transformer with 15°, 30°, 45°, 60°, 75°, and 90° phase shifts are shown in Figs. 1 (a)-(f) solid black curves. The corresponding fractional orders are 0.17, 0.33, 0.5, 0.67, 0.83, and 1, respectively.

To implement the fractional Hilbert transformer, we use a transversal filter approach that offers high performance and reconfigurability in terms of filtering shapes [51-59]. In a transversal filter, discrete frequency samples of the optical signal containing the RF modulation are time delayed, weighted, summed together, and finally detected by a photodetector to generate the RF output. This is essentially equivalent to the filters in digital signal processing (DSP), but implemented by photonic hardware. The transfer function of a photonic transversal filter is given by:

$$H(\omega) = \sum_{n=0}^{N-1} h(n) e^{-j\omega nT} \quad (3)$$

where $\omega$ is the angular frequency of the input RF signal which also corresponds to the RF frequency of the vector network analyser (VNA), the number of taps (n) corresponds to the number of lines from the Kerr comb, $h(n)$ is the discrete impulse response representing the tap coefficient of the $n^{th}$ taps, $N$ is the number of taps, $T$ is the time delay between adjacent taps, and the free spectral range in the RF (FSR$_{RF}$) of the transversal filter is $FSR_{RF} = 1/T$. A fractional Hilbert transformer with arbitrary fractional order can be realized by properly setting the tap coefficients, i.e., $h(n)$, $n = 0, 1, …, N$-1. The corresponding amplitude response of the fractional Hilbert transformers with 15°, 30°, 45°, 60°, 75°, and 90° phase shifts as functions of the number of taps are shown in Figs. 1 (a)-(f). As the number of taps increases, the discrepancy between the ideal amplitude response of the fractional Hilbert transversal filters and that of the transversal filter decreases for all six fractional orders. To quantify these discrepancies, we show the RMSEs in Fig. 2, where we see that they are inversely proportional to the number of taps, as expected. In particular, we note that when the number of taps increases, the RMSE decreases dramatically for small numbers of taps and then decreases more gradually as the number of taps becomes larger.

## III. EXPERIMENTAL SETUP

Figure 3 shows a schematic diagram of the fractional Hilbert transformer. It consists of two main modules: (i) the micro-comb generation module based on a nonlinear micro-ring resonator (MRR), and (ii) a transversal filter module for the reconfigurable FHT with tunable fractional order.

In the first module, optical microcombs were generated in an integrated MRR fabricated on a high-index doped silica glass platform using CMOS-compatible fabrication process [30-36]. First, high-index (n=~1.7 at 1550 nm) doped silica glass films were deposited using plasma enhanced chemical vapour deposition, then patterned by deep ultraviolet photolithography and processed via reactive ion etching to form waveguides with exceptionally low surface roughness. Finally, silica (n = ~1.44 at 1550 nm) was deposited as an upper cladding. The advantages of our platform for optical micro-comb generation include ultra-low linear loss (~0.06 dB·cm$^{-1}$), a moderate nonlinear parameter (~233 W$^{-1}$·km$^{-1}$), and in particular a negligible nonlinear loss up to extremely high intensities (~25 GW·cm$^{-2}$). Due to the ultra-low loss of our platform, the MRR features narrow resonance linewidths of 130MHz FWHM [60, 61], corresponding to a quality factor of ~1.5 million. After packaging the device with fiber pigtails, the through-port insertion loss was ~1 dB, assisted by on-chip mode converters. The radius of the MRR was ~592 μm, corresponding to an optical free spectral range of ~0.4 nm or 50GHz. To obtain optimal parametric gain, the MRR was designed to feature anomalous dispersion in the C-band. A continuous-wave pump at a wavelength of λ=1550.500 nm was amplified by an erbium-doped fibre amplifier and the polarization was adjusted via a polarization controller to optimize the power coupled into the MRR. When the pump wavelength was swept across one of the

MRR's resonances with the pump power high enough to provide sufficient parametric gain, optical parametric oscillation occurred, ultimately generating Kerr optical combs with a spacing equal to the free spectral range of the MRR (Figure 4(c)). Soliton crystal combs with distinctive 'fingerprint' optical spectra were observed, which arose from spectral interference between tightly packaged solitons in the cavity [29]. We note that the soliton crystal micro-combs were generated through adiabatic pump wavelength sweeping – easily achievable via manual tuning.

In the second module, the generated soliton crystal comb was spectrally shaped via two-stage optical shapers (Finisar, WaveShaper 4000S) to enable a larger dynamic range of loss control as well as a higher shaping accuracy (brought about by the intrinsic properties of the WaveShaper). The micro-comb was first shaped to reduce the initial power difference between the comb lines to < 5 dB. In these experiments we used a maximum of 17 taps, or wavelengths, from the micro-comb. The space between the 8th and 9th taps and 9th and 10th taps was 0.8 nm, corresponding to $\Delta\lambda = 0.4 \times 2 = 0.8$ nm, while the spacing between the remaining taps was twice this, at 1.6 nm. This was done in order to increase the Nyquist zone of ~ 50GHz. Next, the shaped comb lines were fed into an electro-optical intensity modulator and then transmitted through a spool of standard single mode fibre (length: L=2.1 km, dispersion: β = ~17.4 ps / nm / km) to provide a wavelength dependent time delay $\tau = L \times \beta \times \Delta\lambda = $~29.4 ps, corresponding to an FSR$_{RF}$ of $1/2\tau = $~17GHz. The operational bandwidth of the fractional Hilbert transformer was mainly determined by the FSR$_{RF}$. An increased operation bandwidth can be obtained by employing a dispersive fibre with a shorter length.

TABLE I
PERFORMANCE OF THE FRACTIONAL HILBERT TRANSFORMER

| Order | Phase shift | Lower cutoff frequency | Upper cutoff frequency | Octaves | Passband ripple | Temporal pulse RMSE |
|-------|-------------|------------------------|------------------------|---------|-----------------|---------------------|
| 0.166 | 15° | < 10 MHz * | 17.1 GHz | >10 * | <0.9 dB | 2.73% |
| 0.333 | 30° | < 10 MHz * | 17.1 GHz | >10 * | <1.1 dB | 2.75% |
| 0.5 | 45° | 0.035 GHz | 16.67 GHz | > 8 | <1.7 dB | 2.69% |
| 0.667 | 60° | 0.49 GHz | 16.6 GHz | > 5 | <2.3 dB | 2.69% |
| 0.833 | 75° | 0.48 GHz | 16.5 GHz | > 5 | <2.9 dB | 2.92% |
| 1 | 90° | 0.48 GHz | 16.45 GHz | > 5 | <2.9 dB | 2.85% |

* the theoretical lower limit is zero (DC). The 10MHz listed here is not fundamental but arises from the lower frequency limit of the vector network analyzer

The shaped comb lines were then amplified and accurately shaped by a second WaveShaper according to the designed tap coefficients. A feedback control path was adopted to increase the accuracy of the comb shaping for the second WaveShaper, where the powers of the comb lines from another output port of the WaveShaper were detected by an optical spectrum analyzer and compared with the ideal tap weights to generate feedback error signals for calibration. Finally, the weighted comb lines representing positive and negative taps were separated by the WaveShaper and converted into electronic signals by a balanced photodetector.

## IV. RESULTS AND DISCUSSION

The system RF frequency response was characterized using a vector network analyser (VNA, Agilent MS4644B) to measure the system RF amplitude and phase frequency response. First, the VNA was calibrated using the 9th (t = 0) tap before the measurement and then the signal output $RF_{SIG}$ was measured with the calibrated VNA.

Figure 5(a) and (b) show the experimental measurements of the RF amplitude frequency response and phase response of the fractional Hilbert transformer with a 45-degree phase shift for 5, 9, 13 and 17 taps, respectively. As predicted in Fig. 5 (a), the operation bandwidth increases with the number of taps. Since our device was designed to perform a fractional Hilbert transform for base-band RF signals, the operating frequency range started at 0.48GHz and ended at 16.45GHz. With 17 tap filters, the fractional Hilbert transformer exhibited a -3dB bandwidth extending from 0.48GHz to 16.45GHz, corresponding to more than 5 octaves, with a phase variation of about ±0.07 rad over the 3-dB pass-band. It is possible to further increase this bandwidth further by using more comb lines in the filter, although in these experiments, only the central part of the C band was actually used to realize the filter taps, limited by the capability of the WaveShaper to compensate for second-order dispersion. For a Hilbert transformer with a 90° phase shift, we were able to use more than 40 taps, thereby reducing the RMSE and amplitude ripple within the pass-band [51]. As seen in Fig. 5 (c), the theoretical 3-dB bandwidth increases rapidly with the number of taps up to 17 taps, after which it levels off, so that the benefit of increasing the number of taps is somehow limited in other words, 17 is close to the optimum number of taps.

We also performed system demonstrations of real-time signal fractional Hilbert transforms for Gaussian-like input pulses with a full-width at half-maximum (FWHM) of ~200 ps, generated by an arbitrary waveform generator (AWG, KEYSIGHT M9505A), as shown in Fig. 6.

Since the spectrum of the comb generated by the MRR did not follow the requirements needed to produce the impulse response of a fractional Hilbert transformer, it was necessary to shape the comb in order to achieve the designed positive and negative tap coefficients. The optical spectra in Fig. 7 (i) was measured by an optical spectrum analyzer (OSA, with a resolution of 0.5 nm). A good match between the power of the measured comb lines (red solid lines for the negative tap coefficients and blue lines for the positive tap coefficients) and the calculated ideal taps weights (orange dots for the negative tap coefficients and green dots for the positive tap coefficients) was obtained, indicating that the comb lines were accurately shaped. We note that the spikes observable in Fig.6 in both the experimentally measured and simulated FFT results are real and are a result of the AWG. However, while the AWG did introduce some glitches, overall the signal was very pure and the spikes contained very little energy; hence, they did not significantly affect the measurements.

Figures 7 (ii)-(iii) present the simulated (red dashed curves) and measured (blue solid lines) transmission response magnitude and phase. The normalized frequency and phase response of the fractional Hilbert transformer with tunable orders of 0.166, 0.333, 0.5, 0.667, 0.833, 1, corresponding to phase shifts of 15°, 30°, 45°, 60°, 75°, 90° are shown. The imbalance of the S21 response around the null frequencies is due to high order dispersion and modulator chirp causing variation in the spacing between taps [49]. The amplitude variations are < 3dB between 480MHz and 16.45GHz and the phase variations are about ±0.07 rad within the 3-dB passband.

Fig. 7 (iv) shows the pulsed waveform after processing by the fractional Hilbert transformer. The detailed performance parameters are listed in Table 1. The output waveform was observed by means of a high-speed real-time oscilloscope (KEYSIGHT DSOZ504A Infinium). For comparison, we show the simulated results (Fig. 7 (iv), red dashed curves). For the 0.166, 0.333, 0.5, 0.667, 0.833, and 1 fractional order Hilbert transformers, the calculated RMSEs between the measured and the theoretical curves are ~2.73%, ~2.75%, ~2.69%, ~2.69%, ~2.92%, ~2.85%, respectively. As can be seen, the measured curves closely match the theoretical counterparts, achieving good performance that agrees well with theory.

## V. CONCLUSION

In summary, we propose and demonstrate a continuously tunable (in its order) microwave and RF fractional Hilbert transformer based on an integrated Kerr micro-comb source. The microcomb produced a large number of comb lines via a CMOS-compatible nonlinear MRR that greatly increases the processing bandwidth. By programming and shaping the comb lines according to the calculated tap weights, we successfully demonstrate a fractional Hilbert transformer with tunable fractional orders of 0.17, 0.33, 0.5, 0.67, 0.83, and 1. We characterize the RF amplitude and phase response, obtaining an operation bandwidth of ~16GHz. We perform system demonstrations of the real-time FHT for Gaussian pulse input signals, obtaining good agreement with theory and verifying that this approach is an effective way to implement high-speed reconfigurable fractional Hilbert transformers featuring high processing bandwidths and reconfigurability, for future ultra-high-speed microwave systems.

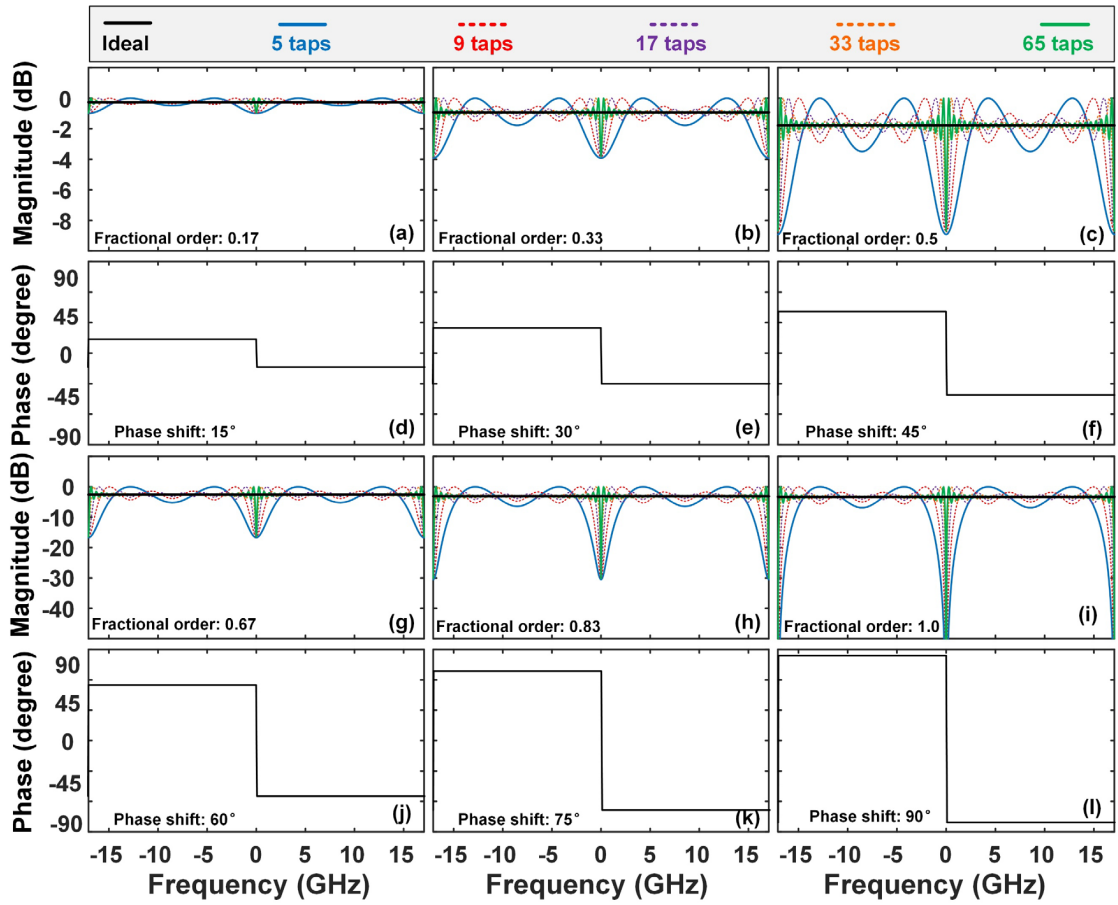

Fig. 1. Theoretical RF amplitude and phase response of FHTs with (a, d) 15°, (b, e) 30°, (c, f) 45°, (g, j) 60°, (h, k) 75°, and (i, l) 90° phase shifts. The amplitude of the fractional Hilbert transformers designed based on Eq. (3) (colour curves) are shown according to the number of taps employed.

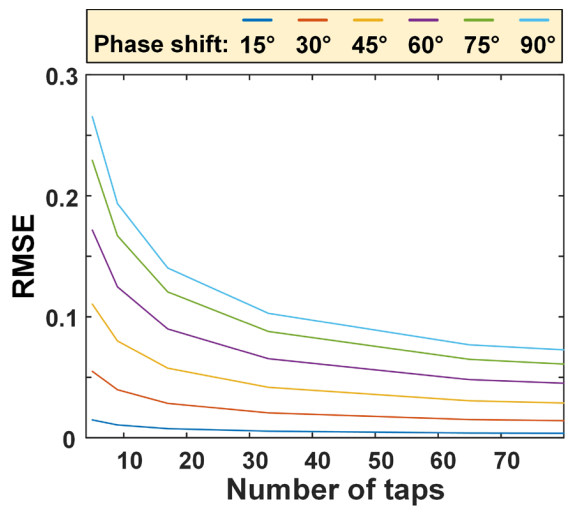

Fig. 2. RMSEs between the calculated and ideal RF amplitude response of the fractional Hilbert transformers with 15°, 30°, 45°, 60°, 75°, and 90° phase shifts as a function of the number of taps.

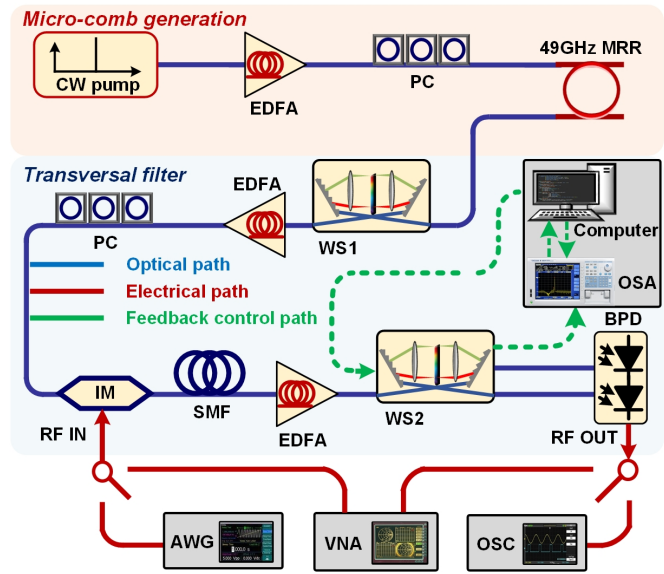

Fig. 3. Schematic diagram of fractional Hilbert transformer based on an integrated Kerr frequency comb source. EDFA: erbium-doped fiber amplifier. PC: polarization controller. MRR: micro-ring resonator. WS: Wave shaper. IM: Intensity modulator. SMF: single mode fiber. OSA: optical spectrum analyzer. BPD: Balanced photodetector. VNA: vector network analyzer. AWG: arbitrary waveform generator. OSC: oscilloscope.

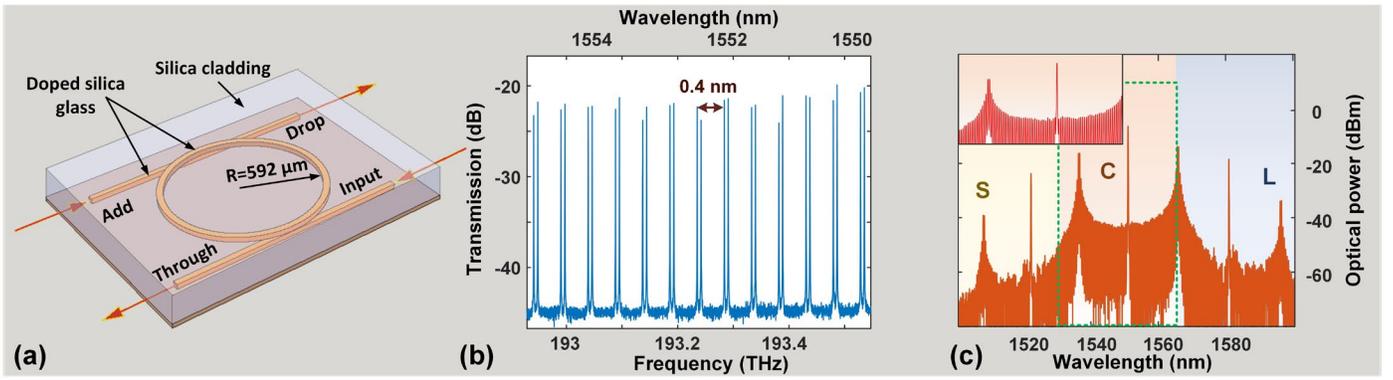

Fig. 4. (a) Schematic illustration of the integrated MRR for generating the Kerr frequency comb. (b) Drop-port optical spectrum of the integrated MRR with a span of 5 nm. (c) Optical spectrum of the generated soliton crystal combs with a 100-nm span. The inset shows a zoom-in spectrum in C-band with a span of ~35 nm.

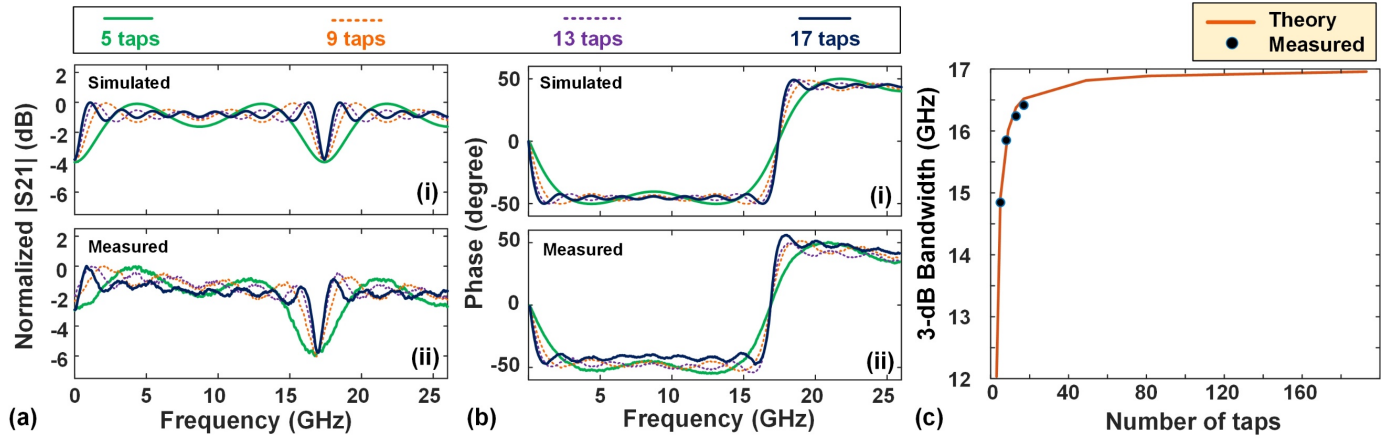

Fig. 5. (a) and (b) Simulated and measured amplitude and phase response for different number of taps for a phase shift of 45-degrees. (c) Simulated and experimental results of 3-dB bandwidth with different taps for a phase shift of 45-degrees.

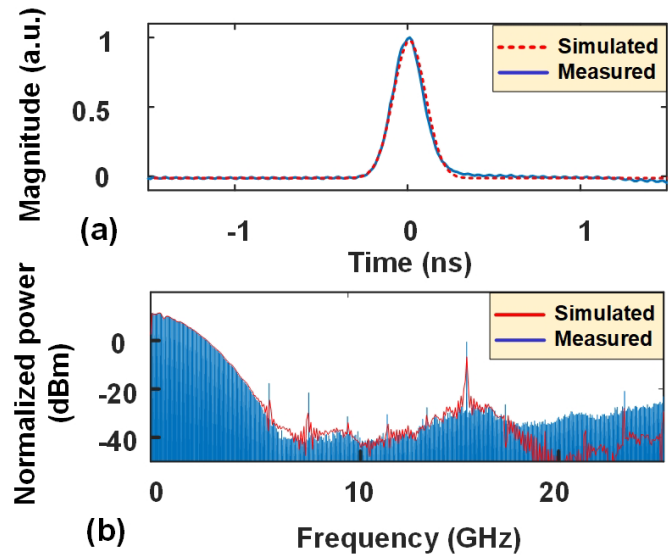

Fig. 6. Measured temporal waveform of a Gaussian input pulse and the simulated and measured FFT of the Gaussian input.

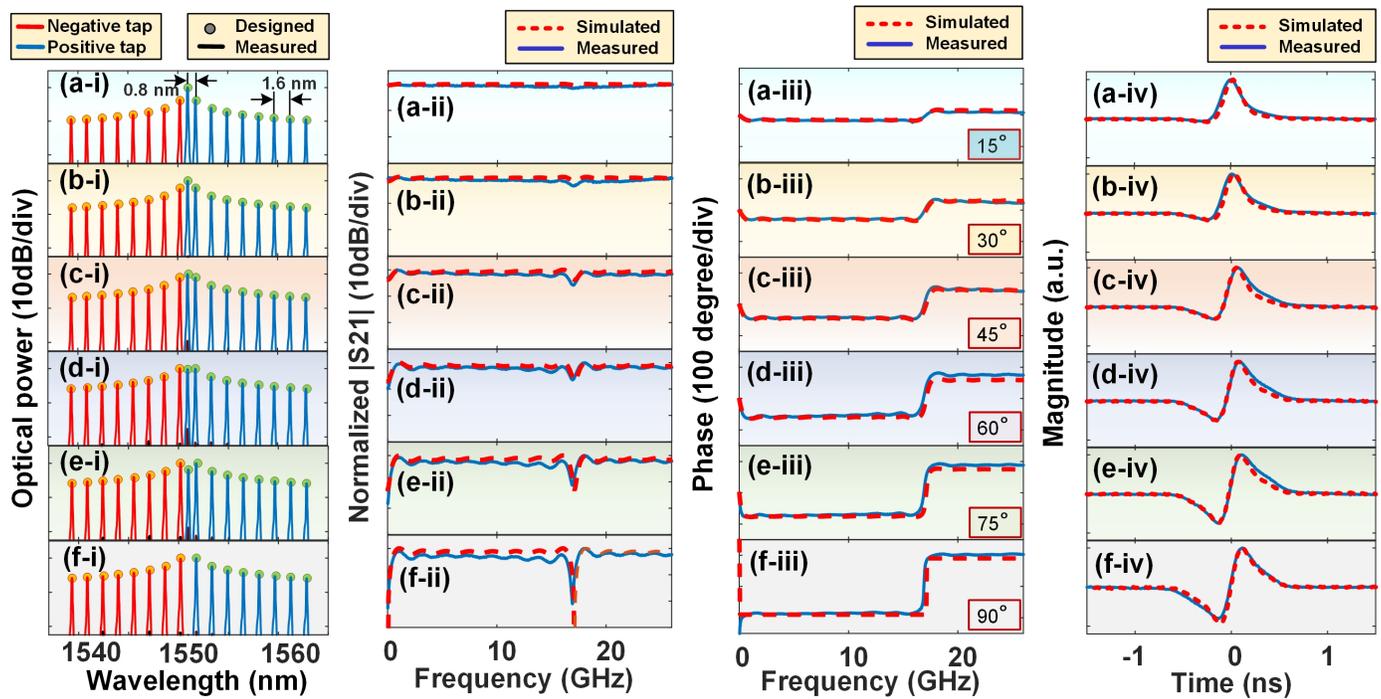

Fig. 7. Simulated (dashed curves) and experimental (solid curves) results of FHT with various phase shifts of (a) 15°, (b) 30°, (c) 45°, (d) 60°, (e) 75°, and (f) 90°. (i) Optical spectra of the shaped micro-comb corresponding with positive and negative tap weights (ii) RF amplitude responses with fractional orders of 0.166, 0.333, 0.5, 0.667, 0.833, and 1. (iii) RF phase responses with phase shifts of 15°, 30°, 45°, 60°, 75° and 90°. (iv) Output temporal intensity waveforms after the FHT.